\documentclass[12pt]{article}
\usepackage{enumerate}
\usepackage{amsfonts}
\usepackage{amsmath}
\usepackage{amssymb}
\usepackage{amsthm}
\usepackage{latexsym}
\usepackage{color}

\def\H{\mathcal{H}}

\def\S{\mathfrak{S}}

\def\B{\mathfrak{B}}

\newcommand{\rank}{\mathrm{rank}}
\newcommand{\id}{\mathrm{Id}}

\newcommand{\shs}{\hspace{1pt}}
\newcounter{defin}  \newcounter{lemma}  \newcounter{theorem}
\newcounter{property} \newcounter{corol}  \newcounter{remark} \newcounter{example}
\newenvironment{lemma}{\par\refstepcounter{lemma}     \textbf{Lemma \thelemma.} }{\rm\par}
\newenvironment{theorem}{\par\refstepcounter{theorem}     \textbf{Theorem \thetheorem.}\ }{\rm\par}
\newenvironment{property}{\par\refstepcounter{property}     \textbf{Proposition \theproperty.}\ }{\rm\par}
\newenvironment{corollary}{\par\refstepcounter{corol}     \textbf{Corollary \thecorol.} }{\rm\par}
\newenvironment{definition}{\par\refstepcounter{defin}     \textbf{Definition \thedefin.}\ }{\rm\par}
\newenvironment{remark}{\par\refstepcounter{remark}     \textbf{Remark \theremark.}}{\rm\par}

\begin{document}

\title{Lower semicontinuity of the entropic disturbance and its applications in quantum information theory}
\author{M.E.~Shirokov, A.S. Holevo \\
Steklov Mathematical Institute, Moscow, Russia}
\date{}
\maketitle
\vspace{-15pt}
\begin{abstract}
We prove that for any infinite-dimensional quantum channel the entropic disturbance (defined as difference between the $\chi$-quantity of a generalized
ensemble and that of the image of the ensemble  under the channel)  is lower
semicontinuous on the  natural set of its definition. We
establish a number of useful corollaries  of this property, in particular, we prove the continuity of the output $\chi\textrm{-}$quantity and the existence of $\chi$-optimal ensemble for any quantum channel  under  the energy-type input constraint.
\end{abstract}
\vspace{-15pt}
%\begin{abstract}

%\end{abstract}
\tableofcontents

\section{Introduction}

Study of various entropic characteristics of a quantum channel is a
significant mathematical problem of quantum information science. Of special
importance are continuity properties as they are related to
robustness and stability of the entropic characteristics with respect to
small perturbations of a state and of a channel. In the present paper we consider the entropic disturbance, defined as difference between
the $\chi $-quantity of a generalized ensemble and that of the image of
the ensemble under an infinite-dimensional quantum channel. This quantity is discussed in \cite{ED-1}, its operational meaning is discovered in \cite{ED-2}. We prove that the entropic disturbance is
lower semicontinuous  in the weak convergence topology on the set of generalized ensembles on which it is
correctly defined (Theorem \ref{chi-loss-ls}) and
establish a number of useful corollaries of this property, in particular,  we prove the continuity of the output $\chi\textrm{-}$quantity and the existence of $\chi $%
-optimal ensemble for any infinite-dimensional quantum channel under the
energy-type input constraint (a problem raised
in \cite{H-Sh-2}).

\section{Notations and preliminaries}

Let $\mathcal{H}$ be a separable Hilbert space, $\mathfrak{B}(\mathcal{H})$
-- the algebra of all bounded operators\ and $\mathfrak{T}(\mathcal{H})$ --
the Banach space of all trace-class operators in $\mathcal{H}$. Let $%
\mathfrak{T}_{+}(\mathcal{H})$ be the cone of positive operators in $%
\mathfrak{T}(\mathcal{H})$ and $\mathfrak{S}(\mathcal{H})$ -- the convex set
of density operators i.e. operators in $\mathfrak{T}_{+}(\mathcal{H})$ with
unit trace, describing \textit{quantum states} \cite{H-SCI,N&Ch}.
Trace-class operators will be usually denoted by the Greek letters $\rho $, $%
\sigma $, $\omega $, ...

We denote by $I_{\mathcal{H}}$ the unit operator in a Hilbert space $%
\mathcal{H}$ and by $\mathrm{Id}_{\mathcal{H}}$ the identity transformation
of the Banach space $\mathfrak{T}(\mathcal{H})$.

A finite or countable collection $\{\rho _{i}\}$ of states with a
probability distribution $\{\pi _{i}\}$ is called \textit{ensemble} and
denoted $\{\pi _{i},\rho _{i}\}$. The state $\bar{\rho}\doteq \sum_{i}\pi
_{i}\rho _{i}$ is called the \emph{average state} of the ensemble.
\smallskip We will also use the notion of \textit{generalized ensemble} as
Borel probability measure on the set of quantum states, so that previously
defined ensembles correspond to discrete probability measures. We denote by $%
\mathcal{P}(\mathcal{H})$ the set of all Borel probability measures on $%
\mathfrak{S}(\mathcal{H})$ equipped with the topology of weak convergence
\cite{Bil,H-Sh-2}. The set $\mathcal{P}(\mathcal{H})$ is a complete
separable metric space \cite{Par}. The average state of a generalized
ensemble $\mu \in \mathcal{P}(\mathcal{H})$ is the barycenter of the measure
$\mu $ defined by the Bochner integral
\begin{equation*}
\bar{\rho}(\mu )=\int_{\mathfrak{S}(\mathcal{H})}\rho \mu (d\rho ).
\end{equation*}

We will use the following compactness criterion for subsets of $\mathcal{P}(%
\mathcal{H})$ \cite[Prop.2]{H-Sh-2}:

\smallskip

\begin{property}
\label{comp-crit} \emph{A closed subset $\mathcal{P}_0$ of $\,\mathcal{P}(%
\mathcal{H})$ is compact if and only if the set $\{\bar{\rho}(\mu)\,|\,\mu\in%
\mathcal{P}_0\}$ is a compact subset of $\,\mathfrak{S}(\mathcal{H})$.}
\end{property}

\smallskip

The \emph{von Neumann entropy} of a quantum state $\rho \in \mathfrak{S}(%
\mathcal{H})$ is defined as $H(\rho )=\mathrm{Tr}\eta (\rho )$, where $\eta
(x)=-x\log x$  for $x>0$ and $\eta (0)=0$. It is a nonnegative, concave and lower semicontinuous function on the set
$\mathfrak{S}(\mathcal{H})$ \cite{L-2,O&P,W}.

The \emph{quantum relative entropy} of states $\rho $ and $\sigma $
in $\mathfrak{T}_{+}(\mathcal{H})$ is defined as follows (cf.\cite{L-2})
\begin{equation*}
H(\rho \,\Vert \hspace{1pt}\sigma )=\sum_{i=1}^{+\infty }\langle i|\,\rho
\log \rho -\rho \log \sigma\,|i\rangle,
\end{equation*}%
where $\{|i\rangle \}_{i=1}^{+\infty }$ is the orthonormal basis of
eigenvectors of the state $\rho $, if $\mathrm{supp}\rho \subseteq \mathrm{supp}\hspace{1pt}\sigma $ and $H(\rho \,\Vert \sigma )=+\infty $ otherwise.\footnote{Throughout the paper we use the Dirac notations, see e.g. \cite{H-SCI,N&Ch}, in which
an orthonormal set of vectors is conventionally denoted as $\{|i\rangle
\}_{i\in I}$, where $I=\{1,2,...,n\}$ or $I=\mathbb{N}$.}

If quantum systems $A$ and $B$ are described by Hilbert spaces $\mathcal{H}%
_{A}$ and $\mathcal{H}_{B}$ then the composite system $AB$ is described by
the tensor product of these spaces $\mathcal{H}_{AB}\doteq \mathcal{H}%
_{A}\otimes \mathcal{H}_{B}$. For a state $\omega _{AB}\in \mathfrak{S}(%
\mathcal{H}_{AB})$, the partial states are $\omega _{A}=\mathrm{Tr}_{%
\mathcal{H}_{B}}\omega _{AB}$ and $\omega _{B}=\mathrm{Tr}_{\mathcal{H}%
_{A}}\omega _{AB}$.\smallskip

The \emph{quantum mutual information} $\,$ of an infinite-dimensional
composite quantum system in the state $\,\omega _{AB}$ is defined as (cf.%
\cite{L-mi})
\begin{equation*}
I(A\!:\!B)_{\omega }=H(\omega _{AB}\hspace{1pt}\Vert \hspace{1pt}\omega
_{A}\otimes \omega _{B})=H(\omega _{A})+H(\omega _{B})-H(\omega _{AB}),
\end{equation*}%
where the second formula is valid if $H(\omega _{AB})<+\infty $. It is well
known that
\begin{equation}
I(A\!:\!B)_{\omega }\leq 2\min \left\{ H(\omega _{A}),H(\omega _{B})\right\}
\label{MI-UB}
\end{equation}%
for any state $\omega _{AB}$ \cite{MI-B,Wilde}.\smallskip

The \emph{Holevo quantity} ($\chi \textrm{-}$\textit{quantity}, for short) of a
generalized ensemble $\mu \in \mathcal{P}(\mathcal{H})$ is defined as (cf.
\cite{H-Sh-2})
\begin{equation}
\chi (\mu )=\int H(\rho\shs \|\shs \bar{\rho}(\mu ))\mu (d\rho )=H(\bar{\rho}(\mu
))-\int H(\rho )\mu (d\rho ),  \label{chi-q-d+}
\end{equation}%
where the second formula is valid under the condition $H(\bar{\rho}(\mu
))<+\infty $.\smallskip\ For a discrete ensemble of states $\,\{\pi
_{i},\rho _{i}\}$ it is equal to
\begin{equation}
\chi (\{\pi _{i},\rho _{i}\})=\sum_{i}\pi _{i}H(\rho _{i}\Vert \bar{\rho})=H(%
\bar{\rho})-\sum_{i}\pi _{i}H(\rho _{i}),  \label{chi-q-d}
\end{equation}%
where the second formula is valid if $H(\bar{\rho})<+\infty $.

A \emph{quantum operation} $\,\Phi$ from a system $A$ to a system $B$ is a
completely positive trace non-increasing linear map $\mathfrak{T}(\mathcal{H}%
_A)\rightarrow\mathfrak{T}(\mathcal{H}_B)$, where $\mathcal{H}_A$ and $%
\mathcal{H}_B$ are Hilbert spaces associated with the systems $A$ and $B$.
In this case we write $\Phi : A\rightarrow B$. A
trace preserving quantum operation is called \emph{quantum channel} \cite%
{H-SCI,N&Ch}.

For any quantum channel $\,\Phi :A\rightarrow B\,$ Stinespring's theorem
(see \cite{St}) implies existence of a Hilbert space $\mathcal{H}_{E}$
(environment) and an isometry $V:\mathcal{H}_{A}\rightarrow \mathcal{H}%
_{B}\otimes \mathcal{H}_{E}$ such that
\begin{equation}
\Phi (\rho )=\mathrm{Tr}_{E}V\rho V^{\ast },\quad \rho \in \mathfrak{T}(%
\mathcal{H}_{A}).  \label{St-rep}
\end{equation}
The minimal dimension of $\mathcal{H}_{E}$ is called  the \emph{Choi rank} of $\hspace{1pt}\Phi$.
The quantum channel  $\widehat{\Phi}: A\rightarrow E$,
\begin{equation}
\widehat{\Phi}(\rho)=\mathrm{Tr}_{B}V\rho V^{\ast }  \label{c-channel}
\end{equation}
is called \emph{complementary} to the channel $\Phi $ \cite[Ch.6]{H-SCI}.\smallskip

Throughout the paper we use the following simple fact:

\begin{remark}
\label{a-ch} There exists a sequence of channels $\Lambda _{n}:A\rightarrow A
$ strongly converging\footnote{This means that $\,\lim_{n\rightarrow \infty }\Lambda _{n}(\rho )=\rho\,$ for
any $\rho \in \mathfrak{S}(\mathcal{H}_{A})$ \cite{AQC}.} to the identity channel $\id_A$ such that
$\Lambda _{n}(\rho )\in\S(\mathcal{H}_{A}^{n})$ for
all $\rho \in \mathfrak{S}(\mathcal{H}_{A})$, where $\mathcal{H}_{A}^{n}$ is a finite-dimensional subspace of $\mathcal{H}_{A}$ for each $n$.
\end{remark}

Such a sequence can be constructed by using any sequence\textit{\ }$\{P_{n}\}$%
\textit{\ }of finite-rank projectors strongly converging to the unit operator%
\textit{\ }$I_{A}$\textit{\ }as follows\textit{\ }%
\begin{equation*}
\Lambda _{n}(\rho )=P_{n}\rho P_{n}+\sigma \mathrm{Tr}(I_{A}-P_{n})\rho ,
\end{equation*}%
where $\sigma $ is a fixed state.\smallskip

For an ensemble $\mu \in \mathcal{P}(\mathcal{H}_{A})$ its image $\Phi (\mu
) $ under a quantum channel $\Phi :A\rightarrow B\,$ is defined as the
ensemble in $\mathcal{P}(\mathcal{H}_{B})$ corresponding to the measure $\mu
\circ \Phi ^{-1}$ on $\mathfrak{S}(\mathcal{H}_{B})$, i.e. $\,\Phi (\mu )[%
\mathfrak{S}_{B}]=\mu[\Phi ^{-1}(\mathfrak{S}_{B})]\,$ for any Borel subset $%
\mathfrak{S}_{B}\subseteq \mathfrak{S}(\mathcal{H}_{B})$, where $\Phi ^{-1}(%
\mathfrak{S}_{B})$ is the pre-image of $\mathfrak{S}_{B}$ under the map $%
\Phi $. If $\mu =\{\pi _{i},\rho _{i}\}$ then $\Phi (\mu )=\{\pi _{i},\Phi
(\rho _{i})\}$.\smallskip

We will use the following continuity condition for the output $\chi \textrm{-}$%
quantity $\chi (\Phi (\mu ))$ \cite[Cor.1]{AQC}. \smallskip

\begin{property}
\label{mu-cont} \emph{Let $\,\Phi:A\rightarrow B\,$ be an arbitrary quantum
channel. The function $\,\mu\mapsto\chi(\Phi(\mu))$ is continuous on a subset $\,\mathcal{P}_0$ of $\,\mathcal{P}(\mathcal{H}_A)$ if the function $\,\mu\mapsto
H(\Phi(\bar{\rho}(\mu)))$ is continuous on $\,\mathcal{P}_0$.}
\end{property}

\smallskip

\begin{remark}
\label{cont}  We will say
that \emph{local continuity of a function $f$ implies local continuity of a
function $g\hspace{1pt}$} if for any sequence $\{x_{k}\}$ converging to $x_{0}$ such that
$\,\lim_{k\rightarrow \infty }f(x_{k})=f(x_{0})\neq \pm \infty\,$ we have
\begin{equation*}
\lim_{k\rightarrow \infty }g(x_{k})=g(x_{0})\neq \pm \infty
\end{equation*}
\end{remark}

\smallskip

We will repeatedly use to the following simple fact. \smallskip

\begin{lemma}
\label{sl} \emph{Let $f_{1},...f_{n}$ be a collection of nonnegative lower
semicontinuous functions on a metric space. Then local continuity of $%
\,\sum_{k=1}^{n}f_{k}$ implies local continuity of all the functions $%
f_{1},...f_{n}$.}
\end{lemma}

\smallskip

\section{Lower semicontinuity of the entropic disturbance}

For a given channel $\,\Phi :A\rightarrow B\,$ and a generalized ensemble $\mu $
the monotonicity of the relative entropy implies
\begin{equation*}
\chi (\Phi (\mu ))\leq \chi (\mu ),
\end{equation*}%
where $\Phi (\mu )$ is the image of the ensemble $\mu $ under action of the
channel $\Phi $. Thus the decrease of $\chi\textrm{-}$quantity
\begin{equation*}
\Delta ^{\!\Phi }\chi (\mu )\doteq \chi (\mu )-\chi (\Phi (\mu ))
\end{equation*}%
(called the entropic disturbance in \cite{ED-1,ED-2}) is a nonnegative function on the set of generalized ensembles with a finite
value of $\chi (\Phi (\mu ))$.\smallskip

\begin{theorem}
\label{chi-loss-ls} \emph{For an arbitrary quantum channel $%
\,\Phi:A\rightarrow B$ the function $\Delta^{\!\Phi}\chi(\mu)$ is lower
semicontinuous on the set $\,\{\hspace{1pt}\mu\in\mathcal{P}(\mathcal{H}%
_A)\,|\,\chi(\Phi(\mu))<+\infty\hspace{1pt}\}$.}\smallskip

\emph{If either the input dimension $d_{A}$ or the Choi rank $d_{E}$ of the
channel $\,\Phi $ is finite then the function $\Delta ^{\!\Phi }\chi (\mu )$
is continuous on the above set and upperbounded by $\,\min \{\log
d_{A},2\log d_{E}\}$.}\smallskip
\end{theorem}

\emph{Proof.} Let $E$ be an environment for $\Phi $ with the minimal
dimensionality $d_{E}$ and $V:\mathcal{H}_{A}\rightarrow \mathcal{H}_{BE}$
be the Stinespring isometry from the representation (\ref{St-rep}). We will
use the identity
\begin{equation}
\chi (\mu )+I(B\!:\!E)_{V\bar{\rho}(\mu )V^{\ast }}=\chi (\Phi (\mu ))+\chi (%
\widehat{\Phi }(\mu ))+\int I(B\!:\!E)_{V\rho V^{\ast }}\mu (d\rho )
\label{b-ident}
\end{equation}%
valid for any $\mu \in \mathcal{P}(\mathcal{H}_{A})$ (with possible values $%
+\infty $ in both sides).

If $\,\dim \mathcal{H}_{A},\dim \mathcal{H}_{B}<+\infty\,$ then the validity of
(\ref{b-ident}) is  verified directly, since in this case $\,I(B\!:\!E)_{V\rho
V^{\ast }}=H(\Phi (\rho ))+H(\widehat{\Phi }(\rho ))-H(\rho )\,$ for any input
state $\rho $. In general case the identity (\ref{b-ident}) can be proved by
approximation (see the Appendix). It implies
\begin{equation}
\chi (\mu )-\chi (\Phi (\mu ))=\chi (\widehat{\Phi }(\mu ))+\int
I(B\!:\!E)_{V\rho V^{\ast }}\mu (d\rho )-I(B\!:\!E)_{V\bar{\rho}(\mu)V^{\ast }}
\label{chi-loss}
\end{equation}%
for any ensemble $\mu $ with finite $\chi (\Phi (\mu ))$ and $I(B\!:\!E)_{V\bar{\rho}(\mu)V^{\ast }}$.

Assume first that the Choi rank $d_{E}\doteq \dim \mathcal{H}_{E}$ of the
channel $\Phi $ is finite. In this case the output entropy of the channel $\,\widehat{\Phi }:A\rightarrow E\,$ is continuous on $\mathfrak{S}(\mathcal{H}_{A})$, so the
function $\chi (\widehat{\Phi }(\mu ))$ is continuous on $\mathcal{P}(%
\mathcal{H}_{A})$ by Proposition \ref{mu-cont}.

The assumption $\dim\mathcal{H}_E<+\infty$ also implies continuity on $%
\mathcal{P}(\mathcal{H}_A)$ of the other terms in the right hand side of (%
\ref{chi-loss}). Indeed, upper bound (\ref{MI-UB}) and Theorem 1A in \cite%
{CMI} show that $\rho\mapsto I(B\!:\!E)_{V\rho V^*}$ is a continuous bounded
function on $\mathfrak{S}(\mathcal{H}_A)$. Hence the continuity of the
second (integral) term in (\ref{chi-loss}) follows from the definition of
the weak convergence topology on $\mathcal{P}(\mathcal{H}_A)$, while the
continuity of the third term follows from continuity of the barycenter map $\,\mu\rightarrow\bar{\rho}(\mu)$.

To prove the upper bound
\begin{equation}
\Delta ^{\!\Phi }\chi (\mu )\doteq \chi (\mu )-\chi (\Phi (\mu ))\leq 2\log
d_{E}  \label{c-l-u-b}
\end{equation}%
note that the triangle inequality $|H(\rho )-H(\Phi (\rho ))|\leq H(\widehat{%
\Phi }(\rho ))\leq \log d_{E}$ (cf.\cite{H-SCI,N&Ch}) directly implies (\ref%
{c-l-u-b}) for any finite ensemble $\mu =\{\pi _{i},\rho _{i}\}$ such that $%
H(\rho _{i})<+\infty $ for all $i$, since in this case
\begin{equation*}
\chi (\mu )-\chi (\Phi (\mu ))=[H(\bar{\rho})-H(\Phi (\bar{\rho}))]-\sum \pi
_{i}[H(\rho _{i})-H(\Phi (\rho _{i}))].
\end{equation*}%
The validity of (\ref{c-l-u-b}) for arbitrary ensemble $\mu $ follows from
the density of the finite ensembles in $\mathcal{P}(\mathcal{H}_{A})$ and
from the continuity of $\Delta ^{\!\Phi }\chi (\mu )$ proved before.

Now we can prove the first assertion of the theorem. By the Stinespring
representation we may assume that $\mathcal{H}_A=\mathcal{H}_{BE}$, $\Phi=%
\mathrm{Tr}_E(\cdot)$ and $\widehat{\Phi}=\mathrm{Tr}_B(\cdot)$. Let $\mu$
be an arbitrary ensemble in $\mathcal{P}(\mathcal{H}_{BE})$. Consider a
sequence of channels $\Lambda^E_n:E\rightarrow E$ strongly converging to the
identity channel $\mathrm{Id}_E$ such that $\,\Lambda^E_n(\mathfrak{S}(%
\mathcal{H}_E))\subseteq\mathfrak{S}(\mathcal{H}^n_E)\,$ for some
finite-dimensional subspace $\mathcal{H}^n_E$ of $\mathcal{H}_E$ (see Remark %
\ref{a-ch}). Let $\mu_n$ be the image of a given ensemble $\mu$ under the
channel $\mathrm{Id}_B\otimes\Lambda^E_n$.

For each $n$ the ensemble $\mu _{n}$ is supported by the subspace $\mathcal{H%
}_{B}\otimes \mathcal{H}_{E}^{n}$. So, speaking about the action of the
channel $\Phi $ on this ensemble we may assume that this channel has finite
Choi rank $\dim \mathcal{H}_{E}^{n}$. Since $\Phi (\mu )=\Phi (\mu _{n})$
for all $n$ and the map $\,\mu \mapsto \mu_{n}\,$ is continuous, the above part
of the proof shows that the function
\begin{equation*}
\mu \,\mapsto \,\chi (\mu _{n})-\chi (\Phi (\mu ))
\end{equation*}%
is continuous on the set of all ensembles $\mu $ with finite $\chi (\Phi
(\mu ))$. Thus, to prove the lower semicontinuty of the function $\mu
\mapsto \chi (\mu )-\chi (\Phi (\mu ))$ on this set it suffices to show that
\begin{equation*}
\chi (\mu _{n})\leq \chi (\mu )\;\textrm{\textrm{ for all }}\;n \quad \mathrm{and}\quad \lim_{n\rightarrow\infty}\chi (\mu
_{n})=\chi (\mu )
\end{equation*}%
for any $\mu \in \mathcal{P}(\mathcal{H}_{BE})$. These relations follow from
the lower semicontinuity of the function $\,\mu \mapsto \chi (\mu)\,$ on the
set $\mathcal{P}(\mathcal{H}_{BE})$ and the monotonicity under the action of
quantum channels.

To complete the proof of the theorem it suffices to say, by Lemma \ref{sl},
that in the case $\,d_{A}\doteq \dim \mathcal{H}_{A}<+\infty\,$ the function $\,\mu \mapsto \chi (\mu)\,$ is continuous on the set $\mathcal{P}(\mathcal{H}%
_{A})$ and is upper bounded by $\log d_{A}$. $\square $ \medskip

Theorem \ref{chi-loss-ls} implies the following condition for local
continuity of the output $\chi\textrm{-}$quantity. \smallskip

\begin{corollary}
\label{chi-loss-c} \emph{For an arbitrary quantum channel $%
\,\Phi:A\rightarrow B$ local continuity of $\,\chi(\mu)$ implies local
continuity of $\,\chi(\Phi(\mu))$, i.e.
\begin{equation*}
\lim_{n\rightarrow\infty}\chi(\mu_n)=\chi(\mu_0)<+\infty
\quad\Rightarrow\quad\lim_{n\rightarrow\infty}\chi(\Phi(\mu_n))=\chi(\Phi(\mu_0))<+\infty
\end{equation*}
for any sequence $\{\mu_n\}\subset\mathcal{P}(\mathcal{H}_{A})$ converging
to an ensemble $\mu_0\in\mathcal{P}(\mathcal{H}_{A})$.}\smallskip
\end{corollary}

\emph{Proof.} By Theorem \ref{chi-loss-ls} and Proposition 1 in \cite{AQC}
all the terms in the equality
\begin{equation*}
\chi(\Phi(\mu))+\Delta^{\!\Phi}\chi(\mu)=\chi(\mu)
\end{equation*}
are lower semicontinuous functions on the set of all ensembles $\mu$ with
finite $\chi(\mu)$. So, the assertion of the corollary follows from Lemma %
\ref{sl}. $\square$\smallskip

Corollary \ref{chi-loss-c} states, briefly speaking, that \emph{local
continuity of the }$\chi\textrm{-}$\emph{quantity is preserved by quantum channels}%
.\smallskip

Combining Corollary \ref{chi-loss-c} and Proposition \ref{mu-cont} we obtain
the following continuity condition for the output $\chi\textrm{-}$quantity, which is
more convenient for applications. \smallskip

\begin{corollary}
\label{chi-loss-c+} \emph{Let $\,\mathfrak{S}_{0}$ be a subset of $\,%
\mathfrak{S}(\mathcal{H}_{A})$ on which the entropy is continuous. Then the
output $\chi\textrm{-}$quantity $\chi (\Phi (\mu ))$ of any quantum channel $\,\Phi
:A\rightarrow B$ is continuous on the set $\,\{\hspace{1pt}\mu \in \mathcal{P%
}(\mathcal{H}_{A})\,|\,\bar{\rho}(\mu )\in \mathfrak{S}_{0}\hspace{1pt}\}$.}
\end{corollary}

\smallskip

In other words, Corollary \ref{chi-loss-c+} states that
\begin{equation*}
\lim_{n\rightarrow\infty}H(\bar{\rho}(\mu_n))=H(\bar{\rho}(\mu_0))<+\infty\quad\Rightarrow\quad%
\lim_{n\rightarrow\infty}\chi(\Phi(\mu_n))=\chi(\Phi(\mu_0))<+\infty
\end{equation*}
for any quantum channel $\Phi:A\rightarrow B$ and any sequence $%
\{\mu_n\}\subset\mathcal{P}(\mathcal{H}_{A})$ converging to an ensemble $%
\mu_0\in\mathcal{P}(\mathcal{H}_{A})$.\smallskip

It is well known (cf.\cite{O&P,W}) that the entropy is continuous on the set
of states $\rho$ satisfying the inequality $\mathrm{Tr} H\rho\leq \mathcal{E}$
provided that the positive operator $H$ satisfies the condition
\begin{equation}  \label{H-cond}
\mathrm{Tr}\hspace{1pt} e^{-\lambda H}<+\infty\;\text{ for all }\;\lambda>0.
\end{equation}
Hence Corollary \ref{chi-loss-c+} implies the following observation which
can be used in continuous variable quantum information theory. \smallskip

\begin{corollary}
\label{chi-loss-c++} \emph{Let $\,\Phi:A\rightarrow B$ be a quantum channel
and $\,\widehat{\Phi}$ its complementary channel. If the Hamiltonian $H_{\!A}$ of
system $A$ satisfies condition (\ref{H-cond}) then the functionals
\begin{equation}  \label{cont-fun}
\mu\,\mapsto\,\chi(\Phi(\mu))\quad \mathit{and}\quad
\mu\,\mapsto\,\chi(\Phi(\mu))-\chi(\widehat{\Phi}(\mu))
\end{equation}
are continuous on the set of all generalized ensembles with bounded average energy
(i.e. on the set $\,\{\mu\in\mathcal{P}(\mathcal{H}_A)\,|\,%
\mathrm{Tr} H_{\!A}\bar{\rho}(\mu)\leq \mathcal{E}\}$).}
\end{corollary}

\medskip

The condition of Corollary \ref{chi-loss-c++} is valid if $A$ is the system
of quantum oscillators and $B$ is any system, in particular, $B=A$ \cite%
{H-SCI,H-c-w-c}. The first functional in (\ref{cont-fun}) is connected to
the unassisted classical capacity of a quantum channel, while the second one
-- to the private classical capacity \cite{H-SCI,Wilde}.

\section{On existence of $\protect\chi$-optimal ensemble for arbitrary
channel}

When we consider transmission of classical information over
infinite-dimensional quantum channel $\,\Phi :A\rightarrow B\,$  we have to impose constraints on states
used for  information encoding to be consistent with the physical
implementation of the process. A typical physically motivated constraint is
the requirement of bounded energy of states used for  information encoding.
This constraint is expressed by the linear inequality\footnote{%
The value $\mathrm{Tr}H_{\!A}\rho $ (finite or infinite) is defined as $\sup_{n}%
\mathrm{Tr}\rho P_{n}H_{\!A}P_{n}$, where $P_{n}$ is the spectral projector of $H_{\!A}$
corresponding to the interval $[0,n]$.}
\begin{equation}
\mathrm{Tr}H_{\!A}\rho \leq \mathcal{E}  \label{lc}
\end{equation}%
where $H_{\!A}$ is a positive self-adjoint operator -- the Hamiltonian of the
input quantum system $A$ and $\mathcal{E}>0$.

The $\chi \textrm{-}$capacity of the
channel $\Phi$ with the constraint (\ref{lc}) can be defined as follows:
\begin{equation}
\bar{C}(\Phi ,H_{\!A},\mathcal{E})=\sup_{\mathrm{Tr}H_{\!A}\bar{\rho}(\mu )\leq\, \mathcal{E}}\chi (\Phi(\mu)),  \label{chi-cap-def}
\end{equation}%
where $\chi (\Phi (\mu ))$ is the output $\chi \textrm{-}$quantity of an ensemble $\mu$ and the supremum is over all ensembles in $\mathcal{P}(\mathcal{H}_{A}) $ with the average state satisfying (\ref{lc}) \cite{H-Sh-2}.

An interesting question concerns attainability of the supremum in (\ref%
{chi-cap-def}). It was formulated in \cite{H-Sh-2} in the following more
general form: under what conditions there is an ensemble $\mu _{\ast }$ such
that
\begin{equation}
\sup_{\bar{\rho}(\mu )\in \mathfrak{S}_{\mathrm{c}}}\chi (\Phi (\mu ))=\chi (\Phi
(\mu _{\ast }))\quad \text{and}\quad \bar{\rho}(\mu _{\ast })\in \mathfrak{S}_{\mathrm{c}}  \label{opt-ens-def}
\end{equation}%
for a given subset $\mathfrak{S}_{\mathrm{c}}$ of $\mathfrak{S}(\mathcal{H}_{A})$.
 Theorem in \cite{H-Sh-2} guarantees the existence of such ensemble
(called $\chi$-optimal) if the set $\mathfrak{S}_{\mathrm{c}}$ is compact and the
output entropy $H(\Phi (\rho ))$ is continuous on $\mathfrak{S}_{\mathrm{c}}$. The
last condition\footnote{The importance of this condition is shown in \cite{H-Sh-2} by proving that  $\chi\textrm{-}$optimal ensemble does not exist for some compact set $\mathfrak{S}_{\mathrm{c}}$ and channel $\Phi$.} is difficult to verify, since in general local continuity of
the entropy is not preserved by quantum channels, i.e. continuity of the
entropy on some set of input states\emph{\ does not imply} continuity (and
even finiteness!) of the output entropy on this set.

The results of Section 3 make it possible to obtain  simpler condition for
existence of $\chi\textrm{-}$optimal ensemble which \emph{does not depend} on a
channel $\Phi$.\smallskip

\begin{property}
\label{opt-ens} \emph{Let $\,\Phi :A\rightarrow B$ be a quantum channel and $%
\mathfrak{S}_{\mathrm{c}}$ be a compact subset of $\,\mathfrak{S}(\mathcal{H}_{A})$. If
the entropy is continuous on $\mathfrak{S}_{\mathrm{c}}$ then (\ref{opt-ens-def})
holds for some ensemble $\mu _{\ast }\in\mathcal{P}(\mathcal{H}_{A})$ supported by pure states.} \smallskip
\end{property}

\emph{Proof.} By Proposition \ref{comp-crit} the set $\mathcal{P}_{\mathrm{c}}\doteq\{%
\hspace{1pt}\mu\in\mathcal{P}(\mathcal{H}_A)\,|\,\bar{\rho}(\mu)\in\mathfrak{%
S}_{\mathrm{c}}\}$ is compact. By Corollary \ref{chi-loss-c+} the function $%
\mu\rightarrow \chi(\Phi(\mu))$ is continuous on the set $\mathcal{P}_{\mathrm{c}}$.
Hence this function achieves its finite maximum on the set $\mathcal{P}_{\mathrm{c}}$,
i.e. (\ref{opt-ens-def}) holds for some ensemble $\mu_0$. By Corollary 6 in
\cite{PPM} there is an ensemble $\mu_*$ supported by pure states such that $%
\mu_*\succ\mu_0$, where $"\succ"$ is the Choquet partial order on the set $%
\mathcal{P}(\mathcal{H}_A)$. Since $\bar{\rho}(\mu_*)=\bar{\rho}(\mu_0)$, the
convexity and lower semicontinuity of the function $\rho\mapsto
H(\Phi(\rho)\|\Phi(\sigma))$ imply, by Lemma 1 in \cite{PPM}, that $%
\chi(\Phi(\mu_*))\geq\chi(\Phi(\mu_0))$. Thus, (\ref{opt-ens-def}) holds for
the ensemble $\mu_*$ as well. $\square$

\medskip

\begin{remark}
\label{opt-ens-r} If the set $\mathfrak{S}_{\mathrm{c}}$ is convex then Proposition 4
in \cite{H-Sh-2} shows that the $\chi\textrm{-}$optimal ensemble $\mu_*$ is characterized by
the property:
\begin{equation*}
\int H(\Phi(\rho)\|\Phi(\bar{\rho}(\mu_*)))\nu(d\rho)\leq \int
H(\Phi(\rho)\|\Phi(\bar{\rho}(\mu_*)))\mu_*(d\rho)=\chi(\Phi(\mu_*))
\end{equation*}
for any ensemble $\nu\in\mathcal{P}(\mathcal{H}_A)$ such that $\bar{\rho}%
(\nu)\in\mathfrak{S}_{\mathrm{c}}$. This property can be considered as a generalization
of the maximal distance property of optimal ensemble for unconstrained
finite-dimensional channels \cite{S&W}. $\square$ \smallskip
\end{remark}

If $\mathfrak{S}_{\mathrm{c}}$ is the set defined by inequality (\ref{lc}) then the
entropy is continuous on $\mathfrak{S}_{\mathrm{c}}$ for all $\,\mathcal{E}>0\,$ if (and only if)
the operator $H_{\!A}$ satisfies the condition (\ref{H-cond}). This condition also
implies compactness of $\mathfrak{S}_{\mathrm{c}}$ (by  Lemma in \cite{H-c-w-c}).
So, we obtain from Proposition \ref{opt-ens} and Remark \ref{opt-ens-r} the
following\smallskip

\begin{corollary}
\label{opt-ens-c} \emph{Let $\,\Phi:A\rightarrow B$ be an arbitrary quantum
channel. If the Hamiltonian $H_{\!A}$ of the system $A$ satisfies condition (\ref%
{H-cond}) then there exists an ensemble $\mu_*\in\mathcal{P}(\mathcal{H}_A)$
supported by pure states such that $\,\mathrm{Tr} H_{\!A}\bar{\rho}(\mu_*)\le \mathcal{E}$,
\begin{equation*}
\chi(\Phi(\mu_*))=\bar{C}(\Phi,H_{\!A},\mathcal{E})\quad\text{and}\quad \int
H(\Phi(\rho)\|\Phi(\bar{\rho}(\mu_*)))\nu(d\rho)\leq\bar{C}(\Phi,H_{\!A},\mathcal{E})
\end{equation*}
for any ensemble $\,\nu\in\mathcal{P}(\mathcal{H}_A)$ such that $\,\mathrm{Tr} H_{\!A}\bar{\rho}(\nu)\le \mathcal{E}$.}
\end{corollary}

\smallskip

If $A$ is the system of quantum oscillators and $B$ is any system, in
particular, $B=A$, then Corollary \ref{opt-ens-c}  proves the existence of $%
\chi\textrm{-}$optimal ensemble supported by pure states for \emph{arbitrary} channel
$\,\Phi:A\rightarrow B\,$ with the energy constraint (\ref{lc}).

\section{On the properties of constrained $\protect\chi$-capacity}

In the analysis of the classical capacity of a quantum channel and of its
relations to other capacities for a given
channel $\Phi :A\rightarrow B$  it is convenient to introduce the function
\begin{equation}
\bar{C}(\Phi ,\rho )\doteq \sup_{\bar{\rho}(\mu )=\rho }\chi (\Phi (\mu ))
\label{chi-fun-def+}
\end{equation}%
on the set $\mathfrak{S}(\mathcal{H}_{A})$ of input states. This function
can be called constrained $\chi\textrm{-}$capacity or simply the $\chi\textrm{-}$function of
the channel $\Phi $ \cite{H-Sh-2,AQC}.\footnote{In \cite{H-Sh-2,AQC} this function is denoted $\chi_{\Phi}(\rho)$.} The $\chi\textrm{-}$capacity of the channel $%
\Phi $ with the linear constraint (\ref{lc}) can be defined via this
function as follows:
\begin{equation*}
\bar{C}(\Phi ,H_{\!A},\mathcal{E})=\sup_{\mathrm{Tr}H_{\!A}\rho \leq\, \mathcal{E}}\bar{C}(\Phi ,\rho ).
\end{equation*}

Note first that Proposition \ref{opt-ens} implies \smallskip

\begin{property}
\label{chi-fun-p+} \emph{For any state $\rho\in\mathfrak{S}(\mathcal{H}_A)$
with finite entropy the supremum in (\ref{chi-fun-def+}) is attained at some
ensemble supported by pure states.}
\end{property}

\smallskip For arbitrary quantum channel $\Phi $ the nonnegative function $%
\rho \mapsto \bar{C}(\Phi ,\rho )$ is concave and lower semicontinuous on $%
\mathfrak{S}(\mathcal{H}_{A})$ \cite{AQC}. By Proposition 5 in \cite{AQC}
continuty of this function on some subset (e.g. converging sequence) of
input states follows from continuity of the output entropy $H(\Phi (\rho ))$
on this set.

The results of Section 3 make it possible to show that continuity of the
function $\rho\mapsto\bar{C}(\Phi,\rho)$ on some subset of input states also
follows from continuity of the input entropy $H(\rho)$ on this set.
\smallskip

\begin{property}
\label{chi-fun-c} \emph{If the entropy is continuous on a subset $\,%
\mathfrak{S}_0$ of $\,\mathfrak{S}(\mathcal{H}_A)$ then the function $%
\,\rho\mapsto\bar{C}(\Phi,\rho)$ is continuous on $\,\mathfrak{S}_0$ for any
channel $\,\Phi$.}
\end{property}

\smallskip \emph{Proof.} By Proposition 4 in \cite{AQC} the function $\rho
\mapsto \bar{C}(\Phi ,\rho )$ is lower semicontinuous on $\mathfrak{S}(%
\mathcal{H}_{A})$. So, it suffices to prove, by Lemma \ref{sl}, that the
function
\begin{equation*}
\rho \,\mapsto \,H(\rho )-\bar{C}(\Phi ,\rho )
\end{equation*}%
is lower semicontinuous on the set of all states $\rho $ with finite $H(\rho)$.

Assume there is a sequence $\{\rho_n\}\subset\mathfrak{S}(\mathcal{H}_A)$
converging to a state $\rho_0$ such that $H(\rho_n)<+\infty$ for
all $\,n\geq0\,$ and  there exists
\begin{equation}  \label{contr}
\lim_{n\rightarrow\infty} \left[H(\rho_n)-\bar{C}(\Phi,\rho_n)\right]< [H(\rho_0)-\bar{C}%
(\Phi,\rho_0)].
\end{equation}
By Proposition \ref{chi-fun-p+}, for each $n$ there exists an ensemble $\mu_n$ in $\mathcal{P}(%
\mathcal{H}_A)$ supported by pure states such that $\bar{C}%
(\Phi,\rho_n)=\chi(\Phi(\mu_n))$ and $\bar{\rho}(\mu_n)=\rho_n$. Since the
set $\{\rho_n\}_{n\geq0}$ is compact, Proposition \ref{comp-crit} in Section
2 implies relative compactness of the sequence $\{\mu_n\}$. So, we may
consider (by passing to a subsequence) that the sequence $\{\mu_n\}$
converges to a particular ensemble $\mu_0\in\mathcal{P}(\mathcal{H}_A)$
supported by pure states. Continuity of the map $\mu\mapsto\bar{\rho}(\mu)$
implies $\bar{\rho}(\mu_0)=\rho_0$. Since $H(\rho_n)=\chi(\mu_n)$ and $%
H(\rho_0)=\chi(\mu_0)$, Theorem \ref{chi-loss-ls} shows that
\begin{equation*}
\begin{array}{rl}
\displaystyle\liminf_{n\rightarrow\infty} \left[H(\rho_n)-\bar{C}(\Phi,\rho_n)\right] \!\!
& =\;\displaystyle\liminf_{n\rightarrow\infty}\left[\chi(\mu_n)-\chi(\Phi(\mu_n))\right] \\\\
\displaystyle & \geq\; \chi(\mu_0)-\chi(\Phi(\mu_0))\geq H(\rho_0)-\bar{C}%
(\Phi,\rho_0),%
\end{array}%
\end{equation*}
where the last inequality follows from definition (\ref{chi-fun-def+}). This
contradicts to (\ref{contr}). $\square$

Note that Proposition \ref{chi-fun-c} implies
\begin{equation}  \label{chi-cap-def++}
\bar{C}(\Phi,H_{\!A},\mathcal{E})=\max_{\mathrm{Tr}H_{\!A}\rho\leq\, \mathcal{E}}\bar{C}(\Phi,\rho),
\end{equation}
provided the operator $H_{\!A}$ satisfies the condition (\ref{H-cond}), since in
this case the set of states such that $\mathrm{Tr}H_{\!A}\rho\leq \mathcal{E}\,$ is compact by
 Lemma in \cite{H-c-w-c}.

\section{On  the gain of entanglement assistance}

The results of Section 4.1 (concerning existence of $\chi\textrm{-}$optimal
ensembles) make it possible to substantially strengthen and simplify the
conditions for equality between the entanglement-assisted classical capacity
and the $\chi\textrm{-}$capacity of an infinite-dimensional quantum channel with
linear constraint presented in \cite[Theorem 2]{EAC-2}, which implies zero
gain in the classical capacity due to entanglement assistance.

 The rate of transmission of classical information over a quantum channel can
be increased by using entangled state as an additional resource. A
detailed description of the corresponding protocol can be found in \cite%
{H-SCI,N&Ch,Wilde}. The ultimate rate of information transmission  in this
protocol is called \emph{entanglement-assisted classical capacity} of a
quantum channel.

If $\,\Phi:A\rightarrow B\,$ is a finite-dimensional quantum channel then the
Bennett-Shor-Smolin-Thaplyal (BSST) theorem~\cite{BSST} gives the following
expression for its entanglement-assisted classical capacity
\begin{equation*}
C_{\mathrm{ea}}(\Phi)=\sup_{\rho \in \mathfrak{S}(\mathcal{H}%
_A)}I(\Phi,\rho),
\end{equation*}
where $\,I(\Phi,\rho)=H(\rho)+H(\Phi(\rho))-H(\widehat{\Phi}(\rho))\,$ is the
quantum mutual information of the channel $\Phi$ at a state $\rho$.
\smallskip

If $\Phi$ is an infinite-dimensional quantum channel then we have to impose
constraint on states used for  information encoding, typically linear
constraint determined by the inequality (\ref{lc}). An operational
definition of the entanglement-assisted classical capacity $C_{\mathrm{ea}%
}(\Phi,H_{\!A},\mathcal{E})$ of an infinite-dimensional quantum channel $\Phi$ with  such a
 linear constraint is given in~\cite{H-c-w-c}, where the generalization
of the BSST theorem is proved under special restrictions on the channel $%
\Phi $ and on the constraint operator $H_{\!A}$. A general version of the BSST
theorem for infinite-dimensional channel with linear constraints without any
simplifying restrictions  which is proved in \cite{EAC-2}, states that
\begin{equation}  \label{eaco}
C_{\mathrm{ea}}(\Phi,H_{\!A},\mathcal{E})=\sup_{\mathrm{Tr}H_{\!A}\rho \leq\,
\mathcal{E}}I(\Phi,\rho)\leq+\infty
\end{equation}
for arbitrary channel $\,\Phi $ and arbitrary constraint operator $H_{\!A}$, where
$I(\Phi,\rho)$ is the quantum mutual information defined by the formula
\begin{equation*}
I(\Phi,\rho) = H\left(\Phi \otimes \mathrm{Id}_{R}
(|\varphi_{\rho}\rangle\langle\varphi_{\rho}|)\hspace{1pt} \|\hspace{1pt}
\Phi (\rho) \otimes \varrho\hspace{1pt}\right),
\end{equation*}
where $|\varphi_{\rho}\rangle$ is a purification of the state $\rho$ in $%
\mathcal{H}_A \otimes \mathcal{H}_R$ and $\varrho=\mathrm{Tr}%
_{A}|\varphi_{\rho}\rangle\langle\varphi_{\rho}|$.\smallskip

In what follows we will assume that the operator $H_{\!A}$ satisfies condition (\ref{H-cond}). So, Corollary \ref{opt-ens-c} guarantees existence of $\chi\textrm{-}$optimal ensemble for arbitrary channel $\Phi$ with the constraint (\ref{lc}), i.e. such ensemble $\mu_*\in\mathcal{P}(\mathcal{H}_A)$ that
\begin{equation}  \label{opt-ens-p}
\bar{C}(\Phi,H_{\!A},\mathcal{E})=\chi(\Phi(\mu_*))\quad\text{and}\quad \mathrm{Tr} H_{\!A}\bar{\rho}(\mu_*)\leq \mathcal{E}.
\end{equation}

We  give the  new conditions for the equality
\begin{equation}
\bar{C}(\Phi,H_{\!A},\mathcal{E})=C_{\mathrm{ea}}(\Phi,H_{\!A},\mathcal{E}).  \label{c-eq}
\end{equation}%
\smallskip

\begin{definition}
\label{c-q-ch-def} A channel $\Phi:A\rightarrow
B$ is called \emph{discrete classical-quantum}
(briefly, \emph{discrete c-q}) channel if it has the representation
\begin{equation}  \label{c-q-ch}
\Phi(\rho)=\sum_k\langle k|\rho|k\rangle\sigma_k,
\end{equation}
where $\{|k\rangle\}$ is an orthonormal basis in $\mathcal{H}_A$ and $%
\{\sigma_k\}$ is a collection of states in $\mathfrak{S}(\mathcal{H}_B)$.
\end{definition}
\smallskip

\begin{definition}
\label{subchannel} Let $\mathcal{H}^0_A$ be a subspace of $\mathcal{H}_A$.
The restriction of a channel $\Phi: A\rightarrow B$ to the subspace  $\mathfrak{T}(\mathcal{H}^0_A)\subset \mathfrak{T}(\mathcal{H}_A)$
is called \emph{subchannel of $\Phi$ corresponding to the subspace $\mathcal{H}^0_A $} and is denoted $\Phi_{\mathcal{H}^0_A}$.
\end{definition}

\smallskip

\begin{definition}
\label{suf-subsp} A subspace $\mathcal{H}^0_A$ of $\mathcal{H}_A$ is called $%
(\bar{C},C_{\mathrm{ea}},H_{\!A},\mathcal{E})$-\emph{sufficient subspace} for a channel $\Phi: A\rightarrow B$ if
\begin{equation*}
\bar{C}(\Phi,H_{\!A},\mathcal{E})=\bar{C}(\Phi_{\mathcal{H}^0_A},H_{\!A},\mathcal{E})\quad \text{and} \quad
C_{\mathrm{ea}}(\Phi,H_{\!A},\mathcal{E})=C_{\mathrm{ea}}(\Phi_{\mathcal{H}^0_A},H_{\!A},\mathcal{E}).
\end{equation*}
\end{definition}

\smallskip

The following theorem is an infinite-dimensional version of Theorem 2 in
\cite{EAC-1}.\smallskip

\begin{theorem}
\label{C-coin} \emph{Let $H_{\!A}$ be a positive operator satisfying condition (%
\ref{H-cond}) and $\mathcal{E}_\mathrm{m}\doteq\inf_{\|\varphi\|=1}\langle\varphi|H_{\!A}|%
\varphi\rangle$ -- the minimal energy level of $H_{\!A}$.}\smallskip

 i) \emph{If $\,\Phi:A\mapsto B$ is an arbitrary channel and (\ref{c-eq})
holds for some $\mathcal{E}>\mathcal{E}_\mathrm{m}$ then there is a $(\bar{C},C_{\mathrm{ea}%
},H_{\!A},\mathcal{E})$-sufficient subspace $\mathcal{H}^0_A$ for $\Phi$ such that $\,\Phi_{%
\mathcal{H}^0_A}$ is a discrete c-q channel (\ref{c-q-ch}) for some basis $%
\,\{|k\rangle\}$ of $\,\mathcal{H}^0_A$. The subspace $\,\mathcal{H}^0_A$
can be defined as the minimal subspace of $\,\H_{\!A}$ containing supports of all ensembles $\mu_*$ satisfying (\ref{opt-ens-p}).}\smallskip

 ii) \emph{If $\,\Phi :A\mapsto B$ is a degradable channel then (\ref{c-eq})
holds for some $\mathcal{E}>\mathcal{E}_{\mathrm{m}}$ if and only if $\,\Phi $ is a discrete c-q
channel (\ref{c-q-ch}), where $\,\{|k\rangle \}$ is the basis of
eigenvectors of $H_{\!A}$ and $\{\sigma_{k}\}$ is a collection of states with
mutually orthogonal supports.}\smallskip
\end{theorem}

\begin{remark}
\label{C-coin-r} The presence of  "$(\bar{C},C_{\mathrm{ea}},H_{\!A},\mathcal{E})$-sufficient
subspace" in Theorem \ref{C-coin} is natural, since the equality $\bar{C}(\Phi,H_{\!A},\mathcal{E})=C_{\mathrm{ea}}(\Phi,H_{\!A},\mathcal{E})$ cannot give information about the
action of the channel $\Phi $ on states absent in the codes
determining $\bar{C}(\Phi,H_{\!A},\mathcal{E})$ and $C_{\mathrm{ea}}(\Phi,H_{\!A},\mathcal{E})$. This is
confirmed by the example of non-entanglement-breaking finite-dimensional
channel $\Phi $ such that $\bar{C}(\Phi)=C_{\mathrm{ea}}(\Phi)$ proposed
in \cite{BSST} and described in \cite[Example 2]{EAC-1}.\smallskip
\end{remark}

\emph{Proof.} i) This assertion follows from Theorem 2 in \cite{EAC-2} and Corollary \ref{opt-ens-c}. It suffices only to note that if (\ref{opt-ens-p}) holds for ensembles $\mu^1_*,\mu^2_*,...$  then it holds for any convex combination $\sum_k p_k\mu^k_*$ of these ensembles (as probability measures).\smallskip

ii) By Corollary \ref{opt-ens-c} the equality $\bar{C}(\Phi,H_{\!A},\mathcal{E})=C_{\mathrm{ea}}(\Phi,H_{\!A},\mathcal{E})$
implies existence of a generalized ensemble with the average
state $\rho_*$ such that $\bar{C}(\Phi,\rho_*)=I(\Phi,\rho_*)=C_{\mathrm{ea}}(\Phi,H_{\!A},\mathcal{E})$ and $\mathrm{Tr} H_{\!A}\rho_*\leq \mathcal{E}$. Since for any degradable
channel $\Phi$ we have $\bar{C}(\Phi,\rho)\leq H(\rho)\leq I(\Phi,\rho)$ for
any state $\rho$, it is easy to see that $\rho_*$ is the state with maximal
entropy under the condition $\mathrm{Tr} H_{\!A}\rho_*\leq \mathcal{E}$, i.e. the Gibbs
state $[\mathrm{Tr} e^{-\lambda^*H_{\!A}}]^{-1}e^{-\lambda^*H_{\!A}}$, where $\lambda^*$
is a solution of the equality $\mathcal{E}\mathrm{Tr} e^{-\lambda^*{H_{\!A}}}=\mathrm{Tr} H_{\!A}
e^{-\lambda H_{\!A}}$. So, $\rho_*$ is a full rank state and Theorem 2 in \cite{EAC-2} shows that $\Phi$ is a discrete c-q channel.

Thus, the  Lemma \ref{C-coin-l} below makes it possible to reduce
assertion ii) to the following observation
\begin{equation*}
\bar{C}(\Pi,H_{\!A},\mathcal{E})=C_{\mathrm{ea}}(\Pi,H_{\!A},\mathcal{E})\quad \Leftrightarrow\quad \langle
k |H_{\!A}| k^{\prime}\rangle=0\text{ for all }\,k\neq k^{\prime},
\end{equation*}
where $\,\Pi(\rho)=\sum_k\langle k|\rho|k\rangle|k\rangle\langle k|\,$ and $\,\mathcal{E}>\mathcal{E}_\mathrm{m}$, proved in \cite[Example 3]{EAC-1}.\smallskip

\begin{lemma}\label{C-coin-l} \cite{EAC-1} \emph{A discrete c-q channel (\ref{c-q-ch}) is
degradable if and only if the collection $\{\sigma_k\}$ consists of states
with mutually orthogonal supports. In this case $\bar{C}(\Phi,H_{\!A},\mathcal{E})=\bar{C}(\Pi,H_{\!A},\mathcal{E})$
and $C_{\mathrm{ea}}(\Phi,H_{\!A},\mathcal{E})=C_{\mathrm{ea}}(\Pi,H_{\!A},\mathcal{E})$ for any
operator $H_{\!A}$ and $\mathcal{E}>0$, where $\Pi(\rho)=\sum_k\langle
k|\rho|k\rangle|k\rangle\langle k|$.}
\end{lemma}

\smallskip

It is easy to show that a channel $\Phi$ has a discrete c-q subchannel
having form (\ref{c-q-ch}) if and only if $\Phi(|k\rangle\langle
k^{\prime}|)=0$ for all $\,k\neq k^{\prime}$. Hence Theorem \ref{C-coin}
implies sufficient conditions for the strict inequality
\begin{equation}  \label{c-ineq}
C_{\mathrm{ea}}(\Phi,H_{\!A},\mathcal{E})>\bar{C}(\Phi,H_{\!A},\mathcal{E}),
\end{equation}
which means that using the entangled state between the input and the output
increases the ultimate speed of information transmission over the channel $\Phi$
and gives a gain in the size of an optimal code:\smallskip

\begin{corollary}
\label{C-coin-c} \emph{Let $H_{\!A}$ be a positive operator satisfying condition (%
\ref{H-cond}) and $\mathcal{E}_\mathrm{m}\doteq\inf_{\|\varphi\|=1}\langle\varphi|H_{\!A}|%
\varphi\rangle$. Then (\ref{c-ineq}) holds  for a channel $\,\Phi$
and $\,\mathcal{E}>\mathcal{E}_\mathrm{m}$ if one of the following condition
is valid:
\begin{itemize}
  \item $\Phi(|\varphi\rangle\langle \psi|)\neq0\,$ for any orthogonal unit vectors $\varphi$ and $\psi$;\footnote{This condition means that $\Phi^*(\B(\H_B))$ is a transitive subspace of $\B(\H_A)$ \cite{TS}.}
  \item $\Phi$ is a degradable channel which is not a discrete c-q channel;
  \item $\Phi$ is a degradable channel and $\,\Phi(|\varphi\rangle\langle \psi|)\neq0\,$ for at least two orthogonal eigenvectors of $\varphi$ and $\psi$ of the operator $H_{\!A}$ corresponding to different eigenvalues;
  \item $\Phi$ is  not a discrete c-q channel and the maximum in (\ref{chi-cap-def++}) is attained at some full rank state.
\end{itemize}}
\end{corollary}

Consider application of Corollary \ref{C-coin-c} to the class of Bosonic
Gaussian channels playing a central role in  the continuous-variable quantum information theory.

Let $\mathcal{H}_{X}$ $(X=A,B,...)$ be the space of irreducible
representation of the Canonical Commutation Relations
\begin{equation*}
W_X(z)W_X(z^{\prime })=\exp \left(-\frac{i}{2}z^{\top }\Delta _{X}z^{\prime
}\right) W_X(z^{\prime }+z)  %\label{CCR}
\end{equation*}
with a symplectic space $(Z_{X},\Delta _{X})$ and the Weyl operators $%
W_{X}(z)$ \cite[Ch.12]{H-SCI}. Denote by $s_X$ the number of modes of the
system $X$, i.e. $2s_X=\dim Z_X$.\smallskip

A Bosonic Gaussian channel $\Phi :\mathfrak{T}(\mathcal{H}_{A})\rightarrow
\mathfrak{T}(\mathcal{H}_{B})$ is defined via the action of its dual $%
\Phi^{\ast }:\mathfrak{B}(\mathcal{H}_{B})\rightarrow \mathfrak{B}(\mathcal{H%
}_{A})$ on the Weyl operators:
\begin{equation*}
\Phi^{\ast}(W_{B}(z))=W_A(Kz)\exp \left[ il^{\top }z-\textstyle\frac{1}{2}
z^{\top }\alpha z\right],\quad z\in Z_B,  %\label{blc}
\end{equation*}
where $K$ is a linear operator $Z_{B}\rightarrow Z_{A}$, $\,l\,$ is a $\,2s_B
$-dimensional real row and $\,\alpha\,$ is a real symmetric $%
\,(2s_B)\times(2s_B)$ matrix satisfying the inequality
\begin{equation*}  %\label{nid}
\alpha \geq \pm \frac{i}{2}\left[ \Delta _{B}-K^{\top }\Delta _{A}K\right].
\end{equation*}

By applying  unitary displacement transformations an arbitrary Gaussian
channel can be transformed to the Gaussian channel with $l=0$ and the same
matrices $K$ and $\alpha$ (such channel is called \emph{centered} and will
be denoted $\Phi_{K,\alpha}$).\smallskip

It follows from Proposition 5 in \cite{EAC-2} that $\Phi_{K,\alpha}$ is a
discrete c-q channel if and only if $K=0$ (i.e. if and only if $%
\Phi_{K,\alpha}$ is a completely depolarizing channel). Proposition 3 in
\cite{BRC} shows that the first condition in Corollary \ref{C-coin-c} holds
if and only if $\mathrm{Ran}K=Z_A$ (i.e. $\mathrm{rank} K=\dim Z_A$). So,
Corollary \ref{C-coin-c} implies the following \smallskip

\begin{corollary}
\label{C-coin-c+} \emph{Let $H_{\!A}$ be a positive operator satisfying condition (%
\ref{H-cond}) and $\mathcal{E}_\mathrm{m}\doteq\inf_{\|\varphi\|=1}\langle\varphi|H_{\!A}|%
\varphi\rangle$. Then (\ref{c-ineq}) holds for the channel $\,\Phi_{K,\alpha}$
and $\,\mathcal{E}>\mathcal{E}_\mathrm{m}$ if one of the following condition is valid:}
\begin{itemize}
  \item \emph{$\mathrm{Ran}K=Z_A$ (i.e.
    $\rank K=\dim Z_A$);}
  \item \emph{$\Phi_{K,\alpha}$ is a degradable channel;}
  \item \emph{$K\neq0$ and the maximum in (\ref{chi-cap-def++}) is attained at some full rank state.}
\end{itemize}
\end{corollary}

The last condition of Corollary \ref{C-coin-c+} holds if $\,\Phi_{K,\alpha}$
is a nontrivial gauge covariant or contravariant channel and $%
H_{\!A}=\sum_{ij}\epsilon_{ij}a^{\dagger}_ia_j$ -- gauge invariant\footnote{The gauge invariance condition for $H_{\!A}$ can be replaced by the requirement
that condition (18) in \cite{H-L} holds as a strict operator inequality.}
Hamiltonian (here $[\epsilon_{ij}]$ -- is a positive matrix), since in this
case the maximum in (\ref{chi-cap-def++}) is attained at a nondegenerate
Gaussian state -- the average state of the $\chi$-optimal ensemble supported by
coherent states \cite{GHP,H-RMS}.

\section{Lower semicontinuity of the coherent information for degradable
channels}

The \emph{coherent information}
\begin{equation}  \label{CI-def}
I_c(\Phi,\rho)\doteq H(\Phi(\rho))-H(\widehat{\Phi}(\rho))
\end{equation}
of a channel $\Phi$ at a state $\rho$ is an important characteristic related
to the quantum capacity of  the channel \cite{H-SCI,N&Ch,Wilde}.

The function $\rho\mapsto I_c(\Phi,\rho)$ is continuous on any set on which
the input entropy $H(\rho)$ is continuous (\cite[Cor.14]{CMI}), but in
general it is not upper or lower semicontinuous on the set of all input
states (where the difference in (\ref{CI-def}) is well defined).

It is known that $I_c(\Phi,\rho)$ is nonnegative for any degradable channel $%
\Phi$ (i.e. such channel that $\widehat{\Phi}=\Theta\circ\Phi$ for some
channel $\Theta:B\rightarrow E$). We will show that in this case $%
I_c(\Phi,\rho)$ is lower semicontinuous as a function of $\rho$. \smallskip

\begin{property}
\label{CI-lsc} \emph{If $\,\Phi:A\rightarrow B$ is a degradable channel then the function
$\rho\mapsto I_c(\Phi,\rho)$ is lower semicontinuous on the set}
$$
\mathfrak{S}_*\doteq\{\hspace{1pt}\rho\in\mathfrak{S}(\mathcal{H}_A)\,|\,H(\widehat{\Phi}(\rho))<+\infty\hspace{1pt}\}.
$$
\end{property}

\smallskip

\emph{Proof.} Since $H(\Phi(\rho))=H(\widehat{\Phi}(\rho))$ for any pure state $\rho$
and $\widehat{\Phi}=\Theta\circ\Phi$, we have
\begin{equation}  \label{CI-rep}
I_c(\Phi,\rho)=\chi(\Phi(\mu))-\chi(\widehat{\Phi}(\mu))=\Delta^{\Theta}%
\chi(\Phi(\mu))
\end{equation}
for any state $\rho\in\S(\H_A)$ with finite $H(\Phi(\rho))$ and any ensemble $\mu\in\mathcal{P}(\mathcal{%
H}_A)$ supported by pure states such that $\bar{\rho}(\mu)=\rho$. By using approximation technique from \cite{AQC} it is easy to show that (\ref{CI-rep}) holds for any state $\rho$ in $\mathfrak{S}_*$ and corresponding ensemble $\mu$.

Let $\{\rho_n\}\subset\mathfrak{S}_*$ be a sequence converging to a state $%
\rho_0\in\mathfrak{S}_*$. Take any sequence $\{\mu_n\}$ of ensembles of pure
states converging to an ensemble $\mu_0$ such that $\bar{\rho}(\mu_n)=\rho_n$
for all $n$ (such sequence can be constructed by using spectral
decompositions of the states $\rho_n$). Then (\ref{CI-rep}) and Theorem \ref%
{chi-loss-ls} imply
\begin{equation*}
\liminf_{n\rightarrow+\infty}
I_c(\Phi,\rho_n)=\liminf_{n\rightarrow+\infty}\Delta^{\Theta}\chi(\Phi(%
\mu_n))\geq \Delta^{\Theta}\chi(\Phi(\mu_0))=I_c(\Phi,\rho_0).\;\;\square
\end{equation*}

\smallskip

\begin{corollary}
\label{CI-lsc-c} \emph{Continuity of the output entropy $H(\Phi(\rho))$ of a
degradable channel $\,\Phi:A\rightarrow B$ on some subset $\,\mathfrak{S}%
_0\subset \mathfrak{S}(\mathcal{H}_A)$ implies continuity of the input
entropy $H(\rho)$ and of the entropy exchange $H(\widehat{\Phi}(\rho))$ on $%
\,\mathfrak{S}_0$.}
\end{corollary}

\smallskip \emph{Proof.} Since
\begin{equation*}
H(\Phi(\rho))=I_c(\Phi,\rho)+H(\widehat{\Phi}(\rho)),
\end{equation*}
the continuity of the entropy exchange $H(\widehat{\Phi}(\rho))$ on $\,%
\mathfrak{S}_0$ follows from Proposition \ref{CI-lsc} and Lemma \ref{sl}.

Now the continuity of the input entropy $H(\rho)$ on $\,\mathfrak{S}_0$
follows from Proposition 9 in \cite{CMI}. $\square$\smallskip

Corollary \ref{CI-lsc-c} shows, in particular, that
\begin{equation*}
\lim_{n\rightarrow\infty}H(\rho_n)\neq H(\rho_0)
\quad\Rightarrow\quad\lim_{n\rightarrow\infty}H(\Phi(\rho_n))\neq
H(\Phi(\rho_0))
\end{equation*}
for a degradable channel $\Phi$ and any  sequence $\{\rho_n\}$ of input states converging to a state $\rho_0$ with finite $H(\Phi(\rho_0))$.
It is easy to see that this implications is not valid in general. It means,
roughly speaking, that \emph{degradable channels preserve local
discontinuity of the input entropy}.

\section*{Appendix: the proof of equality (\protect\ref{b-ident})}

To relax the condition $\dim \mathcal{H}_{A},\dim \mathcal{H}_{B}<+\infty $
consider sequences of channels $\Lambda _{n}^{B}:B\rightarrow B$ and $%
\Lambda _{n}^{E}:E\rightarrow E$ with finite-dimensional outputs strongly
converging to the identity channels $\mathrm{Id}_{B}$ and $\mathrm{Id}_{E}$
(see Remark \ref{a-ch}). Since\break $I(B\!:\!E)_{\Pi _{n}(V\rho V^{\ast
})}=H(\Lambda _{n}^{B}\circ \Phi (\rho ))+H(\Lambda _{n}^{E}\circ \widehat{%
\Phi }(\rho ))-H({\Pi _{n}(V\rho V^{\ast })})$, where $\Pi _{n}\doteq
\Lambda _{n}^{B}\otimes \Lambda _{n}^{E}$, we have
\begin{equation}
\begin{array}{r}
\displaystyle\chi (\Pi _{n}(V\mu V^{\ast }))+I(B\!:\!E)_{\Pi _{n}(V\bar{\rho}(\mu)
V^{\ast })}=\chi (\Lambda _{n}^{B}\circ \Phi (\mu )) \\
\\
+\chi (\Lambda _{n}^{E}\circ \widehat{\Phi }(\mu ))\displaystyle+\int
I(B\!:\!E)_{\Pi _{n}(V\rho V^{\ast })}\mu (d\rho ).%
\end{array}
\label{b-ident+}
\end{equation}%
By using Proposition 4 in \cite{AQC} and the chain rule for the $\chi\textrm{-}$quantity it is easy to show that
\begin{equation*}
\lim_{n\rightarrow\infty}\chi (\Pi _{n}(V\mu V^{\ast }))=\chi (V\mu V^{\ast })=\chi (\mu
)\leq +\infty
\end{equation*}%
and
\begin{equation*}
\lim_{n\rightarrow\infty}\chi (\Lambda_{n}^{B}\circ \Phi (\mu ))=\chi(\Phi (\mu))\leq
+\infty ,\quad \lim_{n\rightarrow\infty}\chi (\Lambda _{n}^{E}\circ \widehat{\Phi }(\mu))=\chi (\widehat{\Phi }(\mu ))\leq +\infty
\end{equation*}%
So, to derive (\ref{b-ident}) from (\ref{b-ident+}) it suffices to show that
\begin{equation}
\lim_{n\rightarrow\infty}I(B\!:\!E)_{\Pi _{n}(V\bar{\rho}(\mu)V^{\ast })}=I(B\!:\!E)_{V\bar{\rho}(\mu)
V^{\ast }}\leq+\infty  \label{l-rel-1}
\end{equation}
and
\begin{equation}
\lim_{n\rightarrow\infty}\int I(B\!:\!E)_{\Pi _{n}(V\rho V^{\ast })}\mu (d\rho )=\int
I(B\!:\!E)_{V\rho V^{\ast }}\mu (d\rho)\leq+\infty.  \label{l-rel-2}
\end{equation}%
 The limit relation (\ref{l-rel-1}) directly follows from the lower semicontinuity of
the quantum mutual information $I(B\!:\!E)$ and its nonincreasing under
action of the local channel $\Pi _{n}$.  The limit relation (\ref{l-rel-2}) can
be rewritten as follows
\begin{equation}
\lim_{n\rightarrow\infty}\int I(B\!:\!E)_{\omega }\nu _{n}(d\omega )=\int I(B\!:\!E)_{\omega}\nu (d\omega )\leq+\infty ,  \label{l-rel-3}
\end{equation}%
where $\nu _{n}$ and $\nu $ are  the images of the ensemble $\mu $ under the
channels $\Pi _{n}(V(\cdot )V^{\ast })$ and $V(\cdot )V^{\ast }$
respectively. By noting that the strong convergence of a sequence of
channels implies uniform convergence of this sequence on compact subsets of
states, it is easy to show the weak convergence of the
sequence $\{\nu_{n}\}$ to the ensemble $\nu $ (see the proof of Lemma 1 in \cite{AQC}). Since the lower semicontinuity and
nonnegativity of the function $\omega \mapsto I(B\!:\!E)_{\omega }$ implies
lower semicontinuity of the functional $\nu \mapsto \int I(B\!:\!E)_{\omega
}\nu (d\omega )$ \cite{Bil}, to prove (\ref{l-rel-3}) (and hence (\ref%
{l-rel-2})) it suffices to note that the nonincreasing of $I(B\!:\!E)$ under
action of the local channel $\Pi _{n}$ implies
\begin{equation*}
\int I(B\!:\!E)_{\omega }\hspace{1pt}\nu _{n}(d\omega )\leq \int
I(B\!:\!E)_{\omega }\hspace{1pt}\nu (d\omega )
\end{equation*}
for all $n$.
\bigskip

{\bf Acknowledgments.} This work is supported by the Russian Science Foundation
under grant 14-21-00162. We are grateful to M.M.Wilde for useful comments concerning the accepted terminology and for relevant references.

\end{document}